\documentclass[prb,american,aps,showpacs,unsortedaddress,twocolumn,10pt]{revtex4-1}
\usepackage{graphics,bm}
\usepackage{amssymb}
\usepackage{epsfig}
\usepackage{epsfig}
\usepackage{epsf}
\usepackage{graphicx}
\usepackage{amsmath}
\usepackage{mathrsfs}
\usepackage{amsfonts}
\usepackage{color}

\makeatletter

\begin{document}

\title{\bf{Thermally Assisted Current-Driven Skyrmion Motion}}
\author{Roberto E. Troncoso$^{1,2,}$}
\email{R.E.TroncosoCona@gmail.com}
\author{Alvaro S. N\'u\~nez$^1$}
\affiliation{(1) Departamento de F\'isica, Facultad de Ciencias F\'isicas y 
Matem\'aticas, Universidad de Chile, Casilla 487-3, Santiago, Chile}
\affiliation{(2) Centro para el Desarrollo de la Nanociencia y la Nanotecnolog\'ia, CEDENNA, Avda. Ecuador 3493, Santiago 9170124, Chile}

\begin{abstract}
We study the behavior of skyrmions in thin films under the action of stochastic torques arising from thermal fluctuations.  We find that the Brownian motion of skyrmions is described by a stochastic Thiele's equation and its corresponding Fokker-Planck equation. The resulting Fokker-Planck equation is recognized as the one for a high-friction Brownian particle which has been studied extensively in different physical contexts. It is shown that thermal fluctuations favor the skyrmion motion allowing a finite mobility even in presence of pinning traps. We calculate explicitly the mobility tensor of skyrmions in linear response to an electric current finding that it increases with temperature.
\end{abstract}


\maketitle
\section{Introduction} 
Skyrmions are topologically protected vortex-like spin structures\cite{Skyrme,Bogdanov,Khawaja,Armaitis, Ezawa}. Recently observed in chiral magnets\cite{Muhlbauer,Jonietz,Neubauer,Munzer,PMilde,Nagao,Seki}, they have received great attention for potential applications in spintronics because it is possible to manipulate their position with low current densities\cite{Fert}. They were  observed in  bulk magnets MnSi\cite{Muhlbauer,Jonietz,Neubauer}, Fe$_{1-x}$Co$_x$Si\cite{Munzer,PMilde,Pfleiderer,Yu2}, Mn$_{1-x}$Fe$_x$Ge\cite{Shibata}, and FeGe\cite{Yu1} using neutron scattering and Lorentz transmission electron microscopy. In these systems, skyrmions with diameters of about a few tens of nanometers were induced by an external magnetic field. Two dimensional atomic-scale magnetic skyrmion lattices have been created in a monoatomic layer of Fe atoms on top of an Ir(111) surface\cite{Heinze}. Due to their topological nature, skyrmions are stable against moderate perturbations.
Numerical simulations have shown that under the influence of spin transfer torques, isolated skyrmions can be created and manipulated\cite{Iwasaki,Iwasaki2}. Two key factors that motivate the study of skyrmions are their small size ({few tens of nanometers}) and  the relatively low currents densities needed in order to drive their motion\cite{Jonietz,Fert}. Those two factors make the skyrmion system a very promising tool for information storage. Experimentally skyrmions in thin films have been observed at low temperatures; however energy calculations predict that isolated skyrmions are expected to be stable even at room temperature\cite{Heinze}. Proper understanding of the brownian motion of skyrmions arising from thermal fluctuations is a very important open issue regarding skyrmion dynamics. It has been proposed by numerical simulations\cite{Kong1}, and experimentally in Ref. [\onlinecite{nagaosa2}], that skyrmions dynamics is activated when exposed to a thermal gradient and also displays a thermal creep motion in a pinning potential in Ref. [\onlinecite{Lin}].

In this paper we study the random motion of magnetic skyrmions arising from thermal fluctuations. Moving as a coherent entity the skyrmion will respond to the thermal fluctuations obeying a stochastic equation of motion. Such equation of motion reduces to the well known Thiele equation in the limit of zero noise. The structure of this paper is the following. In section II, the structure of such an equation is derived and the corresponding Fokker-Planck  equation is presented. To explore the stochastic behavior of the skyrmion we study, in section III, the solutions of the Fokker-Planck equation in the presence of a confining periodic potential in one direction. We solve the Fokker-Planck equation and study the skyrmion mobility and its dependence on temperature.

\section{Stochastic dynamics}
The starting point is the stochastic Landau-Lifschitz-Gilbert (LLG) equation\cite{Brown,Kubo,Ettelaie,Garcia,Heinonen,Safonov,Rossi,Duine} for the magnetization direction ${\bf \Omega}$ that incorporates adiabatic and non-adiabatic\cite{Berger1,Tatara1,Thiaville,Xiao,Zhang,Tatara2} spin-transfer torques\cite{Berger, Slonczewski}, defined as $-{\bf v}_s\cdot{\mathbf{\nabla}}{\bf \Omega}$  and $\beta{\bf v}_s\cdot{\bf \nabla}{\bf \Omega}$  characterized by the dimensionless parameter $\beta$, respectively. Here, ${\bf v}_s=-\left(pa^3/2eM\right){\bf j}$ is the spin-velocity of the conduction electrons, $p$ is the spin-polarization of the electric current, and $e(>0)$ the elementary charge. The stochastic Landau-Lifshitz-Gilbert equation reads
\begin{align}
\left(\frac{\partial}{\partial t}+{\bf v}_s\cdot\nabla\right){\bf \Omega}={\bf \Omega}&\nonumber\times \left({\bf H}_{\text{eff}}+{\bf h}\right)\\
&+\alpha{\bf \Omega}\times\left(\frac{\partial}{\partial t}+\frac{\beta}{\alpha}{\bf v}_s\cdot\nabla\right){\bf \Omega},
\end{align}
where ${\bf H}_{\text{eff}}$ is the effective field and $\alpha$ the Gilbert damping constant. In the above equation ${\bf h}$ is a Gaussian stochastic magnetic field describing the thermal agitation of the magnetization and obeying a zero average and two-point correlations\cite{Brown}
\begin{align}\label{eq:noisecorrelation}
\langle{h}_j({\bf x},t){h}_{j'}({\bf x}',t')\rangle=\sigma a^2\delta({\bf x}-{\bf x}')\delta_{jj'}\delta(t-t'),
\end{align}
with $j,j'$ the cartesian components and $a^2$ the volume of the finite element grid. The strength of the noise is given by $\sigma=2\alpha k_BT/\hbar$, i.e., proportional to the thermal energy $k_BT$  and the Gilbert damping parameter $\alpha$. Note that we are neglecting the influence of current, and their fluctuations,  on damping and fluctuation strength, because these effects are of higher order\cite{Rebei,Foros}. This relation between noise and damping stems from the fluctuation-dissipation theorem \cite{Brown}; thereby we are implicitly assuming a low-energy approximation.

The Landau-Lifshitz-Gilbert equations have solutions that correspond to particle like excitations propagating with a well defined velocity. Examples of such behavior can be found in the dynamics of domain walls\cite{Yamanouchi,Yamaguchi,allwood,hayashi,parkin,alvaro,alvaro2,alvaro3}. 
The description of single-skyrmion dynamics consists of a basic model for a particle-like motion where the  magnetization vector  ${\bf\Omega}$ is parameterized via collective coordinates.  For a single-skyrmion moving rigidly along the trajectory ${\bf x}(t)$ we take as an ansatz the magnetization profile ${\bf \Omega}({\bf r},t)={\bf \Omega}_0{\bf (}{\bf r}-{\bf x}(t){\bf )}$, where ${\bf \Omega}_0$ represents the static skyrmion texture centered at the origin. The skyrmion profile ${\bf \Omega}_0$ is obtained by minimizing the magnetic energy that includes the contributions from exchange energy,  perpendicular anisotropy, and Dzyaloshinskii-Moriya interaction. Such minimization can be accomplished numerically as described  Ref. [\onlinecite{Knoester}]. Plugging-in the ansatz on the LLG equation and integrating over complete space we get for the skyrmion dynamics, characterized by its collective coordinate ${\bf x}(t)$, the stochastic Thiele's equation
\begin{align}\label{eq:stochasticThiele}
a_{\alpha\beta}\dot{x}^{\beta}(t)=-F_{\alpha}({\bf x}(t))+\eta_{\alpha}(t),
\end{align}
with the $2\times 2$-matrix $a_{\alpha\beta}=-\epsilon_{\alpha\gamma\beta}g_{\gamma}+\alpha {\cal D}_{\alpha\beta}$, $\epsilon_{\alpha\beta\gamma}$ is the Levi-Civita symbol, and where summation over repeated indices is assumed.
Clearly in the absence of noise the Eq. (\ref{eq:stochasticThiele}) is reduced to the so called Thiele's equation \cite{Thiele}. The first term of the matrix  $a$ contains the gyromagnetic vector defined by ${g}_{\alpha}=-\frac{1}{2}\epsilon_{\alpha\beta\gamma}G_{\beta\gamma}$, where $G_{\alpha\beta}=\int d{\bf r}\epsilon_{\gamma\delta\epsilon}\Omega^{\gamma}_0\partial_{\alpha}\Omega^{\delta}_0\partial_{\beta}\Omega^{\epsilon}_0$. This term in the equation of motion describes the Magnus force \cite{Jonietz} exerted by flowing electrons. For a single-skyrmion $g_{\alpha}=g\delta_{\alpha z}=4\pi W$ where $W$ is the winding number, or skyrmion charge, that for our case is $W=-1$. On the other hand, the second contribution represents the dissipative force whose components are ${\cal D}_{\alpha\beta}=\int d{\bf r}\partial_{\alpha}\Omega^{\gamma}_0\partial_{\beta}\Omega^{\gamma}_0$, that obeys  for the single-skyrmion case  ${\cal D}_{\alpha\beta}={\cal D}\delta_{\alpha\beta}$ because of the symmetry of the spin configuration.
Explicitly, the stochastic Thiele's equation reads
\begin{eqnarray}\label{eq:stochasticThiele2}
\alpha{\cal D}\dot{x}+g \dot{y}=-F_{x}+\eta_x,\\
-g\dot{x}+\alpha{\cal D} \dot{y}=-F_{y}+\eta_y.
\end{eqnarray}

The drift velocity of the skyrmion in the Langevin's equation [Eq. (\ref{eq:stochasticThiele})] is affected, on one side, by a deterministic force ${\bf F}$ given by 
$F_{\alpha}({\bf x})=\left[\epsilon_{\alpha\beta\gamma}g_{\beta}-\beta {\cal D}\delta_{\alpha\gamma}\right]v^{\gamma}_s-\frac{\partial V[{\bf x}]}{\partial x_{\alpha}},
$ 
that explicitly can be written as:
\begin{eqnarray}\label{eq:deterministicforce}
F_x&=&\left(-g v^y_s-\beta{\cal D}v^x_s\right)-\partial_x V,\\
F_y&=&\left(g v^x_s-\beta{\cal D}v^y_s\right)-\partial_y V, 
\end{eqnarray}
containing both the gyrotropic and dissipative contribution due to the electron's coupling, and also by a force due to the potential $V[{\bf x}]$ from the surrounding environment, e.g. magnetic impurities, local anisotropies or geometric defects. This term, $V[{\bf x}]=V_H[{\bf x}]+V_A[{\bf x}]$,  derives from an inhomogeneous magnetic field ${\bf H}({\bf x})$ coupled to the magnetization of the ferromagnet and a position-dependent perpendicular anisotropy $A({\bf x})$, with $V_H[{\bf x}]=-\frac{1}{\hbar}\int d{\bf r}\Omega_k({\bf r}-{\bf x})H_k({\bf r})$ and $V_A[{\bf x}]=-\frac{1}{\hbar}\int d{\bf r}\Omega^2_z({\bf r}-{\bf x})A({\bf r})$, respectively. In absence of the potential term Eq. (\ref{eq:stochasticThiele}) reduces to the equation obtained in Ref. [\onlinecite{Kong1}].
On the other hand, in Eq. (\ref{eq:stochasticThiele}) there is a stochastic forcing on the skyrmion motion that has a strength of the Gaussian noise $\eta_j$ that turns out to be  
\begin{align}\label{eq:noisecorrelation2}
\langle{\eta}_j(t){\eta}_{j'}(t')\rangle=\sigma a^2{\cal D}\delta_{jj'}\delta(t-t'),
\end{align}
and hence an effective diffusion constant depends not only on the Gilbert damping but on the dissipative force that is parametrized by the dissipative parameter ${\cal D}$ in the ferromagnet. 

In addition, we are interested in the probability distribution $P[{\bf x};t]$ associated with the skyrmion dynamics, which is defined as the probability density that a skyrmion at time $t$
is in the position ${\bf x}$. The probability distribution is thus formally written as $P[{\bf x}
;t]=\langle\delta\left({\bf x}(t)-{\bf x}\right)\rangle$ and the differential equation of motion that it obeys is known as the Fokker-Planck equation and its derivation constitutes a standard issue in stochastic process\cite{Zinn-Justin,Risken}.  
The Fokker-Planck equation associated with the skyrmion motion, derived from  Eq. (\ref{eq:stochasticThiele}), is\begin{align}\label{eq:fokkerplanck}
\frac{\partial}{\partial t}P=a^{-1}_{\alpha\beta}\partial_{\alpha}\left(F_{\beta}P\right)-\frac{\sigma{\cal D}}{2}a^{-1}_{\alpha\beta}\partial^2_{\alpha\beta}P,
\end{align}
where summation over repeated subscripts is understood. The Eq. (\ref{eq:fokkerplanck}) may be written as ${\partial}P/{\partial t}+\partial S_{\alpha}/{\partial x_{\alpha}}=0$ in agreement with the conservation of probability, where we have introduced the definition for the probability current density $S_{\alpha}=-a^{-1}_{\alpha\beta}F_{\beta}P+\frac{\sigma{\cal D}}{2}a^{-1}_{\alpha\beta}\frac{\partial}{\partial X_{\beta}}P$.  At this point we note that Eq. (\ref{eq:fokkerplanck}) involve only derivatives on the position, due to the structure of Eq. (\ref{eq:stochasticThiele}), and hence corresponds to the standard Brownian-motion theory for a high-friction particle \cite{Risken}.  It is indeed linked to the assumption that the skyrmion does not undergo deformations on its size, i.e., a displacement like rigid motion. When this is not so, this typically leads to an effective mass that quantifies such deformation\cite{nagaosa2} or the emission of spin-waves. 

\section{Skyrmion mobility}
Having discussed the derivation of the stochastic skyrmion motion, which is governed by the stochastic Thiele's equation and equivalently its Fokker-Planck equation, we now focus on the average drift motion of the skyrmion and its mobility, and restrict ourselves to the periodic potential case. 

For simplicity we assume a potential varying along the $x$-direction and given by $V(x)=-V_0\cos(2\pi x/\lambda)$, with the assumption that the period of the potential $\lambda$ is greater than the size $l_{sk}$ of the skyrmion {(typically $l_{sk}\sim 20$ nm which is comparable to the experimentally observed sizes in MnSi, that is, $18$ nm\cite{Muhlbauer,Jonietz,Neubauer})}. It is worth mentioning that the deterministic skyrmion motion under periodic potential, i.e., that described by Thiele's equation (without noise), can be solved exactly. This solution shows a pinning effect, in the $x$-direction, arising from the external potential and constraining its motion along the $y$-direction. That effect is only overcome for current densities above the threshold current value $v_c=2\pi\alpha{\cal D}V_0/\lambda(\alpha\beta{\cal D}^2+g^2)$. Hereafter, we show that this extrinsic pinning behavior is reduced due to thermal influence.

The resulting Fokker-Planck equation is recognized as the one for a high-friction Brownian particle in a periodic potential, which has been studied extensively in different physical contexts\cite{Risken}. In the stationary limit, the probability distribution $P[{\bf x},t]$ can be calculated exactly assuming an homogeneous distribution along $y$-direction. The above statement means that $\partial S_y/{\partial y}\equiv 0$, which leads to the total current of probability  to be a constant, and therefore one obtains the well-known result \cite{Risken}
\begin{align}\label{eq:fokkerplanksolution}
P[x]=e^{-{\cal U}(x)/\Theta}\left[N-\gamma S_x/\Theta\int^{x}_0e^{{\cal U}(x')/\Theta}dx'\right],
\end{align}
with the integration constant $N$ that is obtained requiring the normalization of the distribution probability. The parameters in Eq. (\ref{eq:fokkerplanksolution}) are defined as $\Theta=a^2k_BT\alpha^2{\cal D}^2/\hbar$, $\gamma=g^2+\alpha^2{\cal D}^2$, and the total potential ${\cal U}(x)=\alpha{\cal D}V(x)-(\alpha{\cal D}F^{(0)}_x-gF^{(0)}_y)x$, with the superscript indicating the constant part of Eq. (\ref{eq:deterministicforce}).  The $y$-component of the probability current is found to be
$S_y=\frac{g}{\alpha{\cal D}}S_x-\frac{1}{\alpha{\cal D}}F^{(0)}_yP(x)$, where the constant probability current $S_x$ given by 
\begin{widetext}
\begin{align}\label{eq:probabilitycurrent}
S_x=\frac{\frac{\lambda}{2\pi\gamma}\Theta\left(1-e^{-\lambda (\alpha{\cal D}F^{(0)}_x-gF^{(0)}_y)/\Theta}\right)}{\int^{\lambda}_0\int^{\lambda}_0 e^{[{\cal U}(x)-{\cal U}(x')]/\Theta}  dxdx'-\left(1-e^{-\lambda (\alpha{\cal D}F^{(0)}_x-gF^{(0)}_y)/\Theta}\right)\int^{\lambda}_0 \int^{x}_0e^{-[{\cal U}(x)-{\cal U}(x')]/\Theta} dx dx'}.
\end{align}
\end{widetext}
The average drift velocity, determined directly from the probability distribution ($\langle \dot{\bf x}(t)\rangle=\int d{\bf x}\dot{\bf x}(t)P[{\bf x},t]$), can be calculated exactly and it is found to be given in terms of the constant probability current $S_x$ by the relation $u_{\alpha}=(2\pi/\lambda)\int^{\lambda}_0 dxS_{\alpha}$. Explicitly, the components of the mean drift velocity for the skyrmion turns out to be $u_x= 2\pi S_x$ and $u_y=(g/\alpha{\cal D}) u_x-(1/\alpha{\cal D})F^{(0)}_y$. In the zero-temperature limit an extrinsic pinning effect appears due to the motion under the periodic potential. That effect is illustrative if we restrict ourselves to an electric current density along $x$-direction only, in which case the zero-temperature behavior for the drift velocity is $u_x(\Theta\rightarrow 0)=\frac{2\pi\alpha{\cal D}V_0}{\lambda\gamma} \sqrt{(v^x_s/v_c)^2-1}$, where $v_c=2\pi\alpha{\cal D}V_0/\lambda(\alpha\beta{\cal D}^2+g^2)$ corresponds to the threshold current for depinning the skyrmion. Clearly, as $v^x_s<v_c$ the skyrmion is confined on the bottom of some local minima of the potential, and its	 motion, as well as when $v^x_s>v_c$, is dominated by the gyrotropic and dissipative forces. {It is noteworthy that critical current densities to avoid pinning effects in skyrmion dynamics have been studied in similar contexts \cite{Lin,Liu,Everschor} on chiral magnets. In those cases, the pinning physics has been taken into account by adding phenomenological pinning forces due to the presence of impurities. Magnitudes for depinning current densities have been obtained theoretically\cite{Everschor} and experimentally\cite{Schulz} for skyrmion lattice on MnSi.} Next, we discuss the non-zero temperature regime where the extrinsic pinning is absent.

\begin{figure}[ht]
\begin{center}
\includegraphics[width=2.8in]{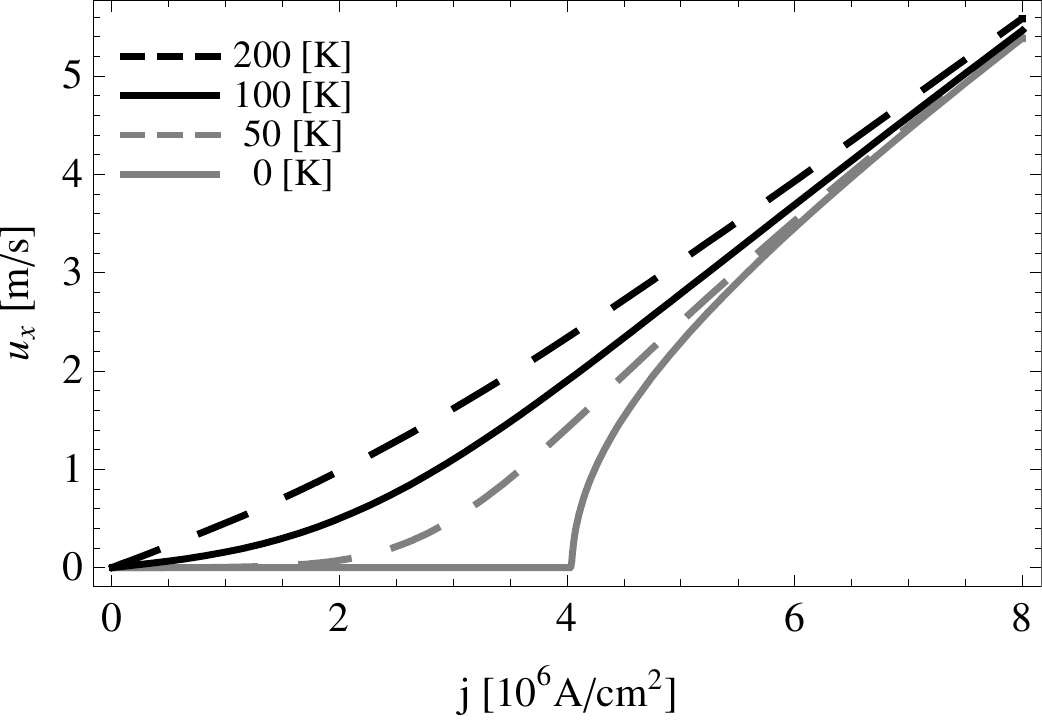}
\includegraphics[width=2.8in]{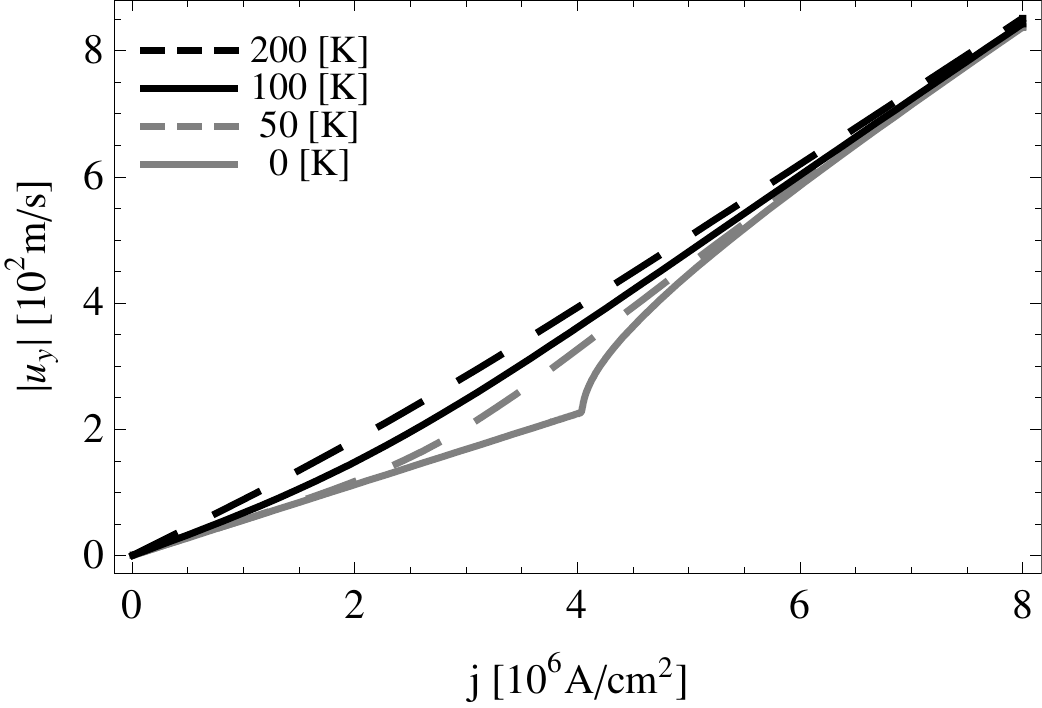}
\caption{Longitudinal (top panel) and transverse (bottom panel) average drift velocity for the thermally-assisted current-induced skyrmion dynamics as a function of electric current and for various temperatures. At zero-temperature the longitudinal skyrmion depinning occur for a critical current with magnitude $j_c\approx 4\times 10^{6} [\text{A}/\text{cm}^2]$. The results are shown for $\alpha=0.01$, $\beta=0.5\alpha$ and a spin-polarization $p=0.2$.}
\label{fig:averagevelocity}
\end{center}
\end{figure}

In Fig. \ref{fig:averagevelocity} the results for the average skyrmion-velocity in the periodic potential are plotted as a function of electric current for various temperatures. Both longitudinal and transversal average  velocity, top and bottom panel respectively, for the thermally-driven skyrmion motion were calculated for Gilbert-damping parameter $\alpha=0.01$, the $\beta$-parameter $\beta=0.5\alpha$, the dissipative force ${\cal D}=5.577\pi$ (from Ref. [\onlinecite{Iwasaki}]). {The nature of the external potential is assumed to be a periodic array of anisotropies arranged along $x$-direction, with a strength of $A=10^{-5}$[eV] and a period of $140[\text{nm}]$. According to the definition of $V_A[{\bf x}]$ we obtain an external potential with a period $\lambda=80$[nm] and $V_0=6Al_{sk}^2/\hbar$}. As we explain above, at zero-temperature the skyrmion displays a locking of the longitudinal motion when $v^x_s<v_c$. A critical current density with magnitude $j_c\approx 4\times 10^{6}[\text{A}/\text{cm}^2]$  enables to the skyrmion avoid the pinning effect. It is a low electric current, however, the longitudinal average velocity $u_x={\cal O}[\text{m/s}]$, which is lesser than that transversal velocities $u_y={\cal O}[10^2\text{m/s}]$. As the temperature is increasing the extrinsic pinning effect is reduced and therefore the skyrmion is capable of moving along the direction of variation of the potential for any value of the current. It is associated with the ability of the skyrmion to tunnel through the potential barrier due to the influence of fluctuating torques. {Despite that the extrinsic pinning stems from a periodic potential, note that the magnitude of our estimated critical current is comparable with values obtained by numerical simulations to depin skyrmions from material disorder\cite{Iwasaki} and triangular notches of stronger perpendicular anisotropy\cite{Fert}.}

It is clearly seen in Fig. \ref{fig:averagevelocity}, and followed from the expression of the mean drift velocity, that at higher temperatures there is a linear dependence of average skyrmion-velocity on current density, indeed we found $u_x\propto -\frac{\alpha\beta{\cal D}^2+g^2}{\alpha^2{\cal D}^2+g^2} v^x_s$ and $u_y\propto-\frac{g}{\alpha{\cal D}}\left(\frac{\alpha\beta{\cal D}^2+g^2}{\alpha^2{\cal D}^2+g^2}+1\right)v^x_s$. When the potential strength is greater, this behavior keeps as we increase the current density. Such tendency of the average velocity is precisely established by means of the calculation of mobility at any temperature. In linear response, i.e. at small electric density currents, the mobility tensor is defined by $\mu_{\alpha\beta}=\partial u_{\alpha}/\partial v^{\beta}_{s}\left.\right|_{v^{\beta}_{s}=0}$. 
Inserting the expression for the probability current Eq. (\ref{eq:probabilitycurrent}) in the result for the average skyrmion-velocity we find that
\begin{align*}
\mu_{xx}(T)&=-\frac{\alpha\beta{\cal D}^2+g^2}{\alpha^2{\cal D}^2+g^2}\left[I_0\left(\frac{V_0\hbar/a^2}{\alpha{\cal D} k_BT}\right)\right]^{-2}\\
\mu_{yx}(T)&=-\frac{g}{\alpha{\cal D}}\frac{\alpha\beta{\cal D}^2+g^2}{\alpha^2{\cal D}^2+g^2} \left[I_0\left(\frac{V_0\hbar/a^2}{\alpha{\cal D} k_BT}\right)\right]^{-2}-\frac{g}{\alpha{\cal D}}
\end{align*}
and
\begin{align*}
\mu_{xy}(T)&=-g{\cal D}\frac{\alpha -\beta}{\alpha^2{\cal D}^2+g^2}\left[I_0\left(\frac{V_0\hbar/a^2}{\alpha{\cal D} k_BT}\right)\right]^{-2}\\
\mu_{yy}(T)&=-\frac{g^2}{\alpha} \frac{\alpha -\beta}{\alpha^2{\cal D}^2+g^2}\left[I_0\left(\frac{V_0\hbar/a^2}{\alpha{\cal D} k_BT}\right)\right]^{-2}+\frac{\beta}{\alpha}
\end{align*}
where $I_0$ is the modified Bessel function of the first kind. 
\begin{figure}[ht]
\begin{center}
\includegraphics[width=3in]{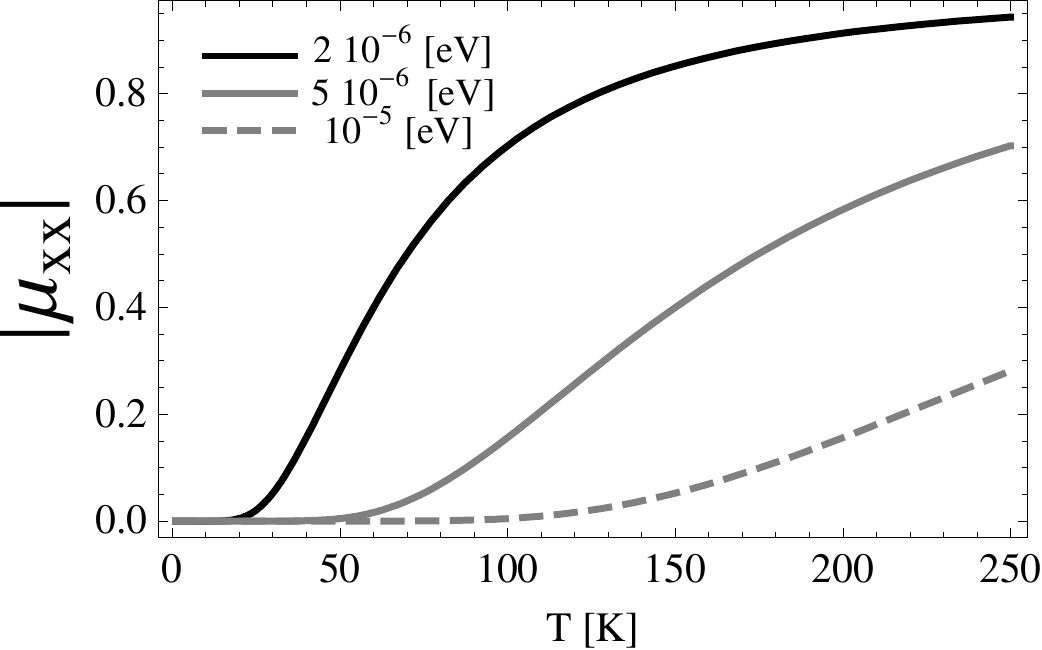}
\caption{Longitudinal skyrmion mobility in linear response as a function of temperature for different values of the strength potential. In the figure is shown the absolute value of the component $\mu_{xx}$ of mobility tensor for the values of the potential $2\times 10^{-6}$[eV], $5\times 10^{-6}$[eV] and $10^{-5}$[eV] represented by the curves black, gray and dashed, respectively. The parameters used to calculate the mobility are the same as in Fig. \ref{fig:averagevelocity}.}
\label{fig:mobility}
\end{center}
\end{figure}
We should also note that taking the high-temperature limit the $xx$- and $yx$-components of the mobility tensor obeys $\mu_{xx}=-\frac{\alpha\beta{\cal D}^2+g^2}{\alpha^2{\cal D}^2+g^2}\approx -1$ and $\mu_{yx}=-\frac{g}{\alpha{\cal D}}\left(\frac{\alpha\beta{\cal D}^2+g^2}{\alpha^2{\cal D}^2+g^2}+1\right)\approx -\frac{2g}{\alpha{\cal D}}$, that recovers our evaluation for the ($v^x_s,u_{\alpha}$) relation. Figure \ref{fig:mobility} shows the curves for the skyrmion mobility (absolute value) as a function of temperature for different strengths of the potential. To link that result with those for the average skyrmion-velocity, we presents a prototype case described by the component $\mu_{xx}$, that quantify the longitudinal skyrmion-motion response under small current densities along $x$-direction. The slow variation of the mobility as we increase the strength of the potential relates to the shift of the critical current density to overcome the pinning of the skyrmion.

\section{Discussion and Conclusions}
We have investigated the mechanisms in which thermal fluctuations influence the current-driven skyrmion motion. Based on stochastic Landau-Lifschitz-Gilbert equation we derived Langevin equation for the skyrmion motion. This equation has the form of a stochastic Thiele's equation, that describes the nonzero-temperature dynamics of a rigid single-skyrmion. From the stochastic Thiele's equation we derived its associated Fokker-Planck equation. By solving explicitly this equation we deduce an exact expression for the average drift velocity of the skyrmion  under a periodic potential.  The longitudinal and transverse mobility of  skyrmions for small applied electric current densities was also determined. 

We find that skyrmions move along the direction of the potential gradient at any nonzero temperature, even for current densities below critical depinning current $j_c\approx 4\times 10^{6} [\text{A}/\text{cm}^2]$, for the parameters used in the main text. This thermally activated depinning of a {current-induced} single-skyrmion is one of our main conclusions, {namely the skyrmion can overcome the energy barrier, and thus, penetrate across the periodic potential due to the thermal agitation}. In addition, thermal fluctuations also affect the transverse motion of the skyrmion increasing its velocity respect to the zero-temperature case.

{In this article a particle-like description has been assumed to treat the skyrmion dynamics, i.e. we consider that its internal structure is rigid. This assumptions is still valid as long as deformations of the skyrmion remain small. The hypothesis of rigid skyrmion is violated by two processes, the deformation of the skyrmion pattern and the emission of spin waves. It is known that the deformation of the skyrmion introduces an inertia term\cite{nagaosa2} (mass-like) into the equations of motion. On the other hand the emission of spin waves is expected to introduce an enhanced damping\cite{Wieser}. To assess the validity of the approximation we use the energy barriers that need to be overcame by the thermal fluctuations in order to excite the internal modes of the skyrmion. Based on the results obtained by Ref. [\onlinecite{Batista}], we then find for MnSi that the minimum thermal energy necessary to excite the internal modes is $k_BT\approx 7$[meV], meaning that the rigid skyrmion-motion hypothesis is valid within the range of temperature $0-80$ [K]. For MnSi we used a strength of exchange and Dzyaloshinskii-Moriya interactions as ${ D}\approx 0.3$[meV], $J\approx3$[meV], respectively (from Ref. [\onlinecite{Batista}]). }

Furthermore, the model implies that at high-temperatures $u_{\alpha}$ varies linearly with current density, with a  proportionality factor that depends on intrinsic parameters only, such as Gilbert damping, skyrmion-charge and dissipative force tensor. This tendency is exhibited by the skyrmion-mobility for high temperatures. 
The role of nonzero temperatures on skyrmion dynamics is expected to be dominant  at low current densities near to the depinning current. Thus, {thermal assisted of a current-driven rigid skyrmion dynamics} could play an important role on the understanding of fundamental physics issues as well as technological applications\cite{nagaosa}. In future work, it might be relevant to investigate a more realistic situation incorporating effects such as, e.g. deformation of the moving skyrmion and its interaction with a random distribution of impurities.

\section{Acknowledgements}
The authors acknowledge funding from Proyecto Fondecyt numbers 11070008 and 1110271, Proyecto Basal FB0807-CEDENNA, Anillo de Ciencia y Tecnonolog\'ia ACT 1117, and by N\'ucleo Cient\'ifico Milenio P06022-F.

\bibliographystyle{elsarticle-num}

\begin{thebibliography}{100}

\bibitem{Skyrme}T. H. R. Skyrme, Nucl. Phys. {\bf 31}, 556 (1962).
\bibitem{Bogdanov}A. N. Bogdanov, U. K. Rossler and A. A. Shestakov  Phys. Rev. E {\bf 67}, 016602 (2003).
\bibitem{Khawaja}U. Al Khawaja and H. T. C. Stoof, Nature {\bf 411} 918 (2001).
\bibitem{Armaitis}J. Armaitis, H. T. C. Stoof, and R. A. Duine, Phys. Rev. Lett. {\bf 110}, 260404 (2013). 
\bibitem{Ezawa} Quantum Hall Effects, Field Theoretical Approach and Related Topics, Zyun Francis Ezawa. World Scientific Publishing Company; 2nd edition (2008).

\bibitem{Muhlbauer}S. Muhlbauer, B. Binz, F. Jonietz, C. Pfleiderer, A. Rosch, A. Neubauer, R. Georgii, P. Boni, 
Science {\bf 323}, 915 (2009).
\bibitem{Jonietz} F. Jonietz , S. Muhlbauer, C. Pfleiderer, A. Neubauer, W. Munzer, A. Bauer, T. Adams, R. Georgii, P. Boni, R. A. Duine, K. Everschor, M. Garst, A. Rosch, Science {\bf 330}, 1648 (2010).
\bibitem{Neubauer}A. Neubauer, C. Pfleiderer, B. Binz, A. Rosch, R. Ritz, P. G. Niklowitz and P. Boni, Phys. Rev. Lett. {\bf 102}, 186602 (2009).

\bibitem{Munzer} W. Munzer, A. Neubauer, T. Adams, S. Muhlbauer, C. Franz, F. Jonietz, R. Georgii, P. Boni, B. Pedersen, M. Schmidt, A. Rosch and C. Pfleiderer, Phys. Rev. B {\bf 81}, 041203(R) (2010).
\bibitem{PMilde}P. Milde, D. Kuhler, J. Seidel, L. M. Eng, A. Bauer, A. Chacon, J. Kindervater, S. Muhlbauer,
 C. Pfleiderer, S. Buhrandt, C. Schutte and A. Rosch, Science {\bf 340}, 1076 (2013).
 \bibitem{Pfleiderer}C. Pfleiderer, T. Adams, A. Bauer, W. Biberacher, B. Binz, F. Birkelbach, P. Boni, C. Franz, R. Georgii, M. Janoschek, F. Jonietz, T. Keller, R. Ritz, S. Muhlbauer, W. Munzer, A. Neubauer, B. Pedersen and A. Rosch, J. Phys. Condens. Matter {\bf 22}, 164207 (2010).
 \bibitem{Yu2} X. Z. Yu, Y. Onose, N. Kanazawa, J. H. Park, J. H. Han, Y. Matsui, N. Nagaosa and Y. Tokura, Nature Materials {\bf 465}, 901 (2010).
\bibitem{Shibata}K. Shibata, X. Z. Yu, T. Hara, D. Morikawa, N. Kanazawa, K. Kimoto, S. Ishiwata, Y. Matsui and Y. Tokura,  Nature Nanotechnology {\bf 8}, 723€" (2013).

\bibitem{Yu1} X. Z. Yu, N. Kanazawa, Y. Onose, K. Kimoto, W. Z. Zhang, S. Ishiwata, Y. Matsui	 and Y. Tokura, Nature Materials {\bf 10}, 106 (2011).

\bibitem{Nagao} M. Nagao, Y. So, H. Yoshida, M. Isobe, T. Hara, K. Ishizuka and K. Kimoto, Nature Nanotechnology {\bf 8}, 325 (2013).
\bibitem{Seki}S. Seki, X. Z. Yu, S. Ishiwata and Y. Tokura, Science {\bf 336}, 198 (2012).
\bibitem{Fert} A. Fert, V. Cros and J. Sampaio, Nature Nanotechnology {\bf 8}, 152 (2013).
\bibitem{Heinze}S. Heinze, K. von Bergmann, M. Menzel, J. Brede, A. Kubetzka, R. Wiesendanger, G. Bihlmayer and S. Blugel, Nature Physics {\bf 7}, 713 (2011).
\bibitem{Iwasaki} J. Iwasaki, M. Mochizuki and N. Nagaosa, Nature Communications {\bf 4}, 1463 (2013).
\bibitem{Iwasaki2}J. Iwasaki, M. Mochizuki and N. Nagaosa, Nature Nanotechnology {\bf 8}, 742€" (2013).

\bibitem{Kong1}L. Kong and J. Zang, Phys. Rev. Lett. {\bf 111}, 067203 (2013).
\bibitem{nagaosa2}M. Mochizuki, X. Z. Yu, S. Seki, N. Kanazawa, W. Koshibae, J. Zang, M. Mostovoy, Y. Tokura and N. Nagaosa, Nature Materials {\bf 13}, 241 (2014).
\bibitem{Lin}S. Z. Lin, C. Reichhardt, C. D. Batista and A. Saxena, Phys. Rev. B {\bf 87}, 214419 (2013).




\bibitem{Brown} W. F. Brown, Jr., Phys. Rev. {\bf 130}, 1677 (1963).
\bibitem{Kubo} R. Kubo and N. Hashitsume, Prog. Theor. Phys. Suppl. {\bf 46}, 210 (1970).
\bibitem{Ettelaie} R. Ettelaie and M.A. Moore, J. Phys. A {\bf 17}, 3505 (1984).
\bibitem{Garcia} J. L. Garc\'ia-Palacios and F. J. Lazaro, Phys. Rev. B {\bf 58},
14937 (1998).
\bibitem{Heinonen} O. G. Heinonen and H. S. Cho, IEEE Transactions on
Magnetics {\bf 40}, 2227 (2004).
\bibitem{Safonov} V. L. Safonov and H. N. Bertram, Phys. Rev. B {\bf 71},
224402 (2005).
\bibitem{Rossi} E. Rossi, O. G. Heinonen, and A.H. MacDonald, Phys.
Rev. B {\bf 72}, 174412 (2005).
\bibitem{Duine}R.A. Duine, A. S. N\'u\~nez, and A.H. MacDonald, Phys. Rev. Lett. {\bf 98} 056605 (2007).


\bibitem{Berger1}L. Berger, J. Appl. Phys. {\bf 49}, 2156 (1978); L. Berger,
J. Appl. Phys. {\bf 71}, 2721 (1992).
\bibitem{Tatara1} G. Tatara and H. Kohno, Phys. Rev. Lett. {\bf 92}, 086601 (2004).
\bibitem{Thiaville} A. Thiaville, Y. Nakatani, J. Miltat and Y. Suzuki, Europhys. Lett. {\bf 69}, 990 (2005).
\bibitem{Xiao} J. Xiao, A. Zangwill and M. D. Stiles, Phys. Rev. B {\bf 73}, 054428 (2006).
\bibitem{Zhang} S. Zhang and Z. Li, Phys. Rev. Lett. {\bf 93}, 127204 (2004).
\bibitem{Tatara2} G. Tatara, H. Kohno, and J. Shibata, J. Phys. Soc. Jpn. {\bf 77}, 031003 (2008).


\bibitem{Berger}L. Berger, J. Appl. Phys. {\bf 55}, 1954 (1984).
\bibitem{Slonczewski}J. C. Slonczewski, J. Magn. Magn. Mater. {\bf 159}, L1 (1996).

\bibitem{Rebei} A. Rebei and M. Simionato, Phys. Rev. B {\bf 71}, 174415
(2005).
\bibitem{Foros} J. Foros, A. Brataas, Y. Tserkovnyak, and G. E. W. Bauer, Phys. Rev. Lett. {\bf 95}, 016601 (2005).

\bibitem{Yamanouchi} M. Yamanouchi, D. Chiba, F. Matsukura, T. Dietl and H. Ohno, Phys. Rev. Lett. {\bf 96}, 096601 (2006).
\bibitem{Yamaguchi} A. Yamaguchi, T. Ono, S. Nasu, K. Miyake, K. Mibu and T. Shinjo, Phys. Rev. Lett. {\bf 92}, 077205 (2004).
\bibitem{allwood} D. A. Allwood, G. Xiong, C. C. Faulkner, D. Atkinson, D. Petit and R. P. Cowburn, Science {\bf 309}, 1688 (2005).
\bibitem{hayashi}M. Hayashi, L. Thomas, R. Moriya, C. Rettner and S. S. P. Parkin, Science {\bf 320}, 209 (2008).
\bibitem{parkin} S. S. P. Parkin, M. Hayashi and L. Thomas, Science {\bf 320}, 190 (2008). 
\bibitem{alvaro} P. Landeros and Alvaro S. N\'u\~nez, J. Appl. Phys. {\bf 108}, 033917 (2010).
\bibitem{alvaro2} J.A. Otalora, J.A. L\'opez-L\'opez, P. Landeros, P. Vargas and A.S. N\'u\~nez, J. Magn. Magn. Mat. {\bf 341}, 86 (2013).
\bibitem{alvaro3} J.A. Otalora, J.A. L\'opez-L\'opez, A.S. N\'u\~nez and P. Landeros,   J. Phys. Cond. Matter {\bf 24}, 436007 (2012).
\bibitem{Knoester}M.E. Knoester, J. Sinova and R.A. Duine,  Phys. Rev. B {\bf 89}, 064425  (2014), A. Bogdanov and A. Hubert, J. Magn. Magn. Mater. {\bf 138}, 255 (1994), N.S. Kiselev, A.N. Bogdanov, R. Shafer and U.K. Rossler, J. Phys. D {\bf 44}, 392001 (2011).

\bibitem{Thiele} A. A. Thiele, Phys. Rev. Lett. {\bf 30}, 230 (1972).
\bibitem{Zinn-Justin}J. Zinn-Justin, Quantum Field Theory and Critical
Phenomena (Oxford, New York, 1989).
\bibitem{Risken} H. Risken, The Fokker-Planck Equation (Springer-Verlag,
Berlin, 1984).

\bibitem{Liu}Ye-Hua Liu and You-Quan Li J. Phys.: Condens. Matter {\bf 25} 076005 (2013).
\bibitem{Everschor}K. Everschor, M. Garst, B. Binz, F. Jonietz, S. Muhlbauer, C. Pfleiderer and A. Rosch, Phys. Rev. B {\bf 86}, 054432 (2012).
\bibitem{Schulz}T. Schulz,	R. Ritz, A. Bauer, M. Halder, M. Wagner, C. Franz, C. Pfleiderer, K. Everschor, M. Garst and A. Rosch, Nature Physics {\bf 8}, 301 (2012).
\bibitem{Wieser}R. Wieser, E. Y. Vedmedenko, and R. Wiesendanger, Phys. Rev. B {\bf 81}, 024405 (2010); Y. Le Maho, Joo-Von Kim, and G. Tatara, Phys. Rev. B {\bf 79}, 174404 (2009).
\bibitem{Batista}S.Z. Lin, C. D. Batista and A. Saxena, Phys. Rev. B {\bf 89}, 024415 (2014).
\bibitem{nagaosa}N. Nagaosa and Y. Tokura Nature Nanotechnology {\bf 8}, 899 (2013).







\end{thebibliography}

\end{document}